\documentclass[prb,twocolumn,superscriptaddress,showpacs]{revtex4}

\usepackage{graphicx}
\usepackage{mathrsfs}
\usepackage{amsmath,amssymb}

\begin{document}

\title{Superconductivity in a  Hubbard-Fr\"ohlich Model and in cuprates.}

\author{T. M. Hardy}
\affiliation{Department of Physics, Loughborough University, Loughborough, LE11 3TU, U.K.}

\author{J. P. Hague}
\affiliation{Department of Physics, Loughborough University, Loughborough, LE11 3TU, U.K.}
\affiliation{Department of Physics and Astronomy, The Open University,  MK7 6AA, U.K.}

\author{J. H. Samson}
\affiliation{Department of Physics, Loughborough University, Loughborough, LE11 3TU, U.K.}

\author{A. S. Alexandrov}
\affiliation{Department of Physics, Loughborough University, Loughborough, LE11 3TU, U.K.}

\begin{abstract}

Using the variational Monte-Carlo method we find that a relatively
weak long-range electron-phonon interaction induces a d-wave
superconducting state of doped Mott-Hubbard insulators and/or
strongly-correlated metals with a condensation energy significantly
larger than can be obtained with Coulomb repulsion only. Moreover,
the superconductivity is shown to exist for infinite on-site Coulomb
repulsion, removing the requirement for additional mechanisms such
as spin fluctuations to mediate d-wave superconductivity. We argue
that the superconducting state is robust with respect to a more
intricate choice of the trial function and  that the true origin of
high-temperature superconductivity lies in a proper combination of
strong electron-electron correlations with poorly screened
Fr\"{o}hlich electron-phonon interaction.


\end{abstract}

\maketitle

It is now over 20 years since the discovery of the first high
temperature superconductor \cite{bed} and yet, despite intensive
effort, the origin of the superconductivity remains fundamentally
unknown, with no widely accepted theory.  The absence of consensus
on the physics of the cuprates and the recent discovery of
iron-based compounds with high transition temperatures has
re-emphasized the importance of understanding the origins of
superconductivity in quasi-two-dimensional materials
\cite{hosono2008}. The canonical BCS-Migdal-Eliashberg theory could
hardly account for well documented non-Fermi-liquid properties of
cuprate superconductors. Moreover, calculations based on the local
density approximation (LDA) often predict negligible electron-phonon
interaction (EPI) insufficient to explain a kink in the
quasiparticle energy dispersion observed by ARPES \cite{manske}.
Hence, it is not surprising that a large number of researchers held
the view that the repulsive Hubbard model would have the essential
physics to account for  the superconducting and non-Fermi-liquid
normal states of cuprates. The idea behind this, originally proposed
by Anderson~\cite{RVBPWA}, is that mobile hole pairs are created via
a strong on-site repulsion, $U$.  Results by Paramekanti et al.
~\cite{Paramekanti} and Yamaji et al.~\cite{Yamaji}, using a
variational Monte Carlo (VMC) simulation with a (projected) BCS-type
trial wave function, appeared to back this up.

However recent studies by Aimi and Imada~\cite{Imada}, using a
sign-problem-free  Gaussian-Basis Monte Carlo (GBMC)
algorithm~\cite{GBMC} have shown that these variational methods, as
well as some other approximations, overestimated the normal state
energy and
 therefore overestimated the condensation energy by at least
an order of magnitude, so that the Hubbard model does not account
for high-temperature superconductivity either. This remarkable
result is in line with earlier numerical studies using the
auxiliary-field quantum (AFQMC) \cite{afqmc} and constrained-path
(CPMC) \cite{cpmc} Monte-Carlo methods, none of which found
superconductivity in the Hubbard model.

On the other hand compelling experimental evidence for a strong EPI
has arrived from isotope effects \cite{zhao}, high resolution angle
resolved photoemission spectroscopies (ARPES) \cite{lanzara}, a
number of optical \cite{mic1,ita,tal}, neutron-scattering
\cite{ega,rez}  and some other spectroscopies of cuprates
\cite{davis}. Here we show that a long-range EPI combined with the
Hubbard $U$ provides sizable superconducting order in doped
Mott-Hubbard insulators and/or strongly-correlated metals. Previous
numerics have shown that d-wave order could exist in basic models of
electron-phonon interactions as a consequence of second order
effects (which look similar to those in repulsive models), with d-
selected over s-wave order by the Hubbard repulsion, which acts as a
kind of filter \cite{hague2006a}. Other work determining the
properties of strongly-coupled polarons and bipolarons have shown
that the long-range discrete Fr\"ohlich EPI in the presence of a
strong Coulomb repulsion, does not lead to the enormous enhancement
of the carrier effective mass characteristic of the Hubbard-Holstein
model (HHM)
\cite{asa,dev,alekor,fehske,bonca,alekor2,jim2,nagaosa,polarons}. It
is our thinking that the special properties of the Fr\"ohlich type
interaction are also likely to be interesting in the weak
electron-phonon coupling regime with intermediate carrier density.
Specifically, the purpose of this study is to determine if $d$-wave
superconductivity is enhanced by a combination of the lighter
carrier masses and propensity to intersite pairing found in models
with long-range interactions and significant, repulsive Hubbard $U$.

Our Hubbard-Fr\"ohlich model (HFM) contains the usual electron
hopping, $t({\bf n})$, and electron-electron on-site repulsive
correlations, $U$. In addition there is a term to describe lattice
vibrations, with frequency $\omega$ and mass $M$, and a term to
describe the long-range interaction of the electrons with  ion
displacements,
\begin{eqnarray}
&&H =  \sum_{{\bf n,n'}, \sigma=\uparrow,\downarrow} t({\bf n-n'})
c^{\dag}_{{\bf n'}\sigma} c_{{\bf n}\sigma}  + U \sum_{\bf n}
\hat{n}_{{\bf n} \uparrow} \hat{n}_{{\bf n} \downarrow} + \cr
 && \sum_{\bf m} \Big[\frac{\hat{P}^2_{\bf m}}{2M} + \frac{
M\omega^2 \xi^2_{\bf m}}{2} \Big] - \sum_{{\bf m, n},\sigma}
f_{\bf m}({\bf n}) c^{\dag}_{{\bf n}\sigma} c_{{\bf n}\sigma}
\xi_{\bf m}. \label{ham}
\end{eqnarray}
Here $c^{\dag}_{{\bf n'}\sigma}$ and $c_{{\bf n}\sigma}$ create and
annihilate the electron with spin $\sigma$ at sites ${\bf n'}$ and
${\bf n}$, respectively, $\hat{P}_{\bf m}= -i \hbar \partial /
\partial \xi_{\bf m}$ is the ion momentum operator at site ${\bf
m}$, and $\xi_{\bf m}$ is the ion displacement. The long-range
Fr\"ohlich EPI is characterized by the force which has the form
\cite{alekor},
\begin{equation}
f_{\bf m}({\bf n}) = \frac{\kappa}{[({\bf m-n})^2/a^2 +
1]^{3/2}}\exp \Big(-\frac{|{\bf m-n}|}{R_{sc}}\Big), \label{force}
\end{equation}
where   $\bf m, n$ are the lattice vectors, $a$ is the lattice
constant, and $R_{sc}$ is the screening radius. That is, the
screened force is the unscreened force multiplied by an exponential
screening factor. The Fr\"ohlich EPI,
Eq.(\ref{force}), routinely neglected in the Hubbard $U$ and $t$--$J$
models of cuprate superconductors \cite{ran} is of the order of 1 eV
as estimated from optical  data \cite{asa}. The force function,
Eq.(\ref{force}), describes the EPI with c-axis polarized optical
phonons, which is poorly screened since the upper limit for the
out-of-plane plasmon frequency in cuprates \cite{mar} is well below
the characteristic frequency of optical phonons.

The carrier mass and the range of the applicability of analytical
weak and strong-coupling expansion
 effectively depend on
the EPI radius. In particular the exact carrier mass calculated with
the Fr\"{o}hlich EPI using the continuous-time QMC algorithm
\cite{alekor} was found to be several orders of magnitude smaller
than in the Holstein model in the relevant region of $\hbar
\omega/t$ ratio. The mass is  well fitted by a single exponent  remarkably
close to those obtained using the Lang-Firsov transformation and
subsequent averaging over phonons, for \emph{any strength} of the
Fr\"ohlich EPI, justifying our use of the Lang--Firsov
transformation in this work. Here we use the transformation
\cite{lan} to integrate out phonons (for technical details see for
example Ref.\cite{polarons}) mapping the electron part of the
transformed Hamiltonian, $\tilde{H}$, to an extended
$\tilde{U}$--$\lambda$ Hubbard model with renormalized hopping
integrals, $\tilde{t}({\bf n})$, a diminished on-site repulsion,
$\tilde{U}$, and a long-range effective attraction, $\lambda W$,
\begin{eqnarray}
\tilde{H} &=&  \sum_{{\bf n \neq n'}, \sigma} \tilde{t}({\bf n-n'})
c^{\dag}_{{\bf n'}\sigma} c_{{\bf n}\sigma}  + \tilde{U} \sum_{\bf
n} \hat{n}_{{\bf n} \uparrow} \hat{n}_{{\bf n} \downarrow} \cr
 &-& \lambda W \sum_{{\bf n \neq n'}, \sigma} \Phi({\bf n-n'})
\hat{n}_{{\bf n'}\sigma} \hat{n}_{{\bf n}\sigma}, \label{renorm}
\end{eqnarray}
Here $\tilde{t}({\bf n})=t({\bf n}) \exp[-g^2({\bf n})]$, $g^2({\bf
n})=[E_p-\lambda W\Phi({\bf n})]/\hbar\omega$, $\Phi({\bf n})
=\kappa^{-2} \sum_{\bf m} f_{\bf m}({\bf 0}) f_{\bf m}({\bf n})$,
$\tilde{U}=U-2E_p$, $W=z\tilde{t}({a})$ is the renormalized
half-bandwidth, $z$ is the coordination number and $\lambda =
\kappa^2/2M\omega^2W$ is proportional  the conventional BCS
electron-phonon coupling constant. The latter is proportional to the
single-particle density of states (DOS) and  could be larger than
$\lambda$ due to the van-Hove singularity of DOS and
correlation-enhanced effective mass of carriers.  The polaronic
shift, $E_p=(\kappa^2/2M\omega^2) \Phi({\bf 0})$, of atomic levels
is included in the chemical potential. In the following only
nearest-neighbor hops are allowed. As shown by Bon\v{c}a and Trugman
\cite{bonca} the physical properties of (bi)polaronic carriers
depend predominantly on the EPI coming from the first two sites in
Eq.(\ref{force}), so that we take $R_{sc}=\infty$ in our
simulations.

When EPI is strong compared with the renormalized kinetic energy,
$\lambda \gg 1$, one can apply the $1/\lambda$ perturbation
expansion reducing the multi-polaron problem to a charged Bose-gas
of small mobile bipolarons \cite{asa,polarons}.  Intermediate and
weak coupling regimes, $\lambda \lesssim 1$, with large $\tilde{U}$
require a variational approach. Here we use a standard VMC method as
for example in Ref.~\onlinecite{Paramekanti} to minimize the energy
$\tilde{H} |\Psi \rangle / | \Psi \rangle $. The difference is that
we make an additional measurement of the long-range attraction in
the extended Hubbard Hamiltonian, Eq.(\ref{renorm}).

A BCS type trial wave-function is used, which has the form
\begin{equation}
|\Psi_T\rangle= P_N P_G \prod_{\bf k} (u_{\bf k} + v_{\bf k}
c^{\dag}_{{\bf k} \uparrow} c^{\dag}_{-{\bf k} \downarrow}) | 0
\rangle,\label{trial}
\end{equation}
 where $P_N = \delta_{\sum_i \hat{n}_i,N}$ projects onto a fixed
particle number state and $P_G = \prod_{\bf n} [1 -
(1-g)\hat{n}_{{\bf n}\uparrow}\hat{n}_{{\bf n}\downarrow}]$ projects
out double occupancy~\cite{Yamaji}. Our variational parameters are
$0 \leqslant g \leqslant 1$ of the Gutzwiller double occupancy
projection operator, $P_G$, and the chemical potential, $\mu$, that
enters through the kinetic energy, $\xi_{\bf k}\equiv \tilde{t}(\cos
k_x+\cos k_y) -\mu$, via $u^2_{\bf k} = (1 + \xi_{\bf k} /
\sqrt{\xi^2_{\bf k} + \Delta_{\bf k}^2})/2$ and $v^2_{\bf k} = (1 -
\xi_{\bf k} / \sqrt{\xi^2_{\bf k} + \Delta_{\bf k}^2})/2$. We then
vary the parameters to minimize the energy for values of the
electron density $\rho$, $\tilde{U}$, superconducting order
parameter $\Delta_{\bf k}$ and $\lambda$ keeping $\tilde{t} = 1$. We
place 42 spin-up and 42 spin-down electrons on a 10$\times$10 2D
square lattice to simulate optimal doping, keeping the electron
density fixed at $\rho = 0.84$. Two cases are investigated: a) an
on-site repulsion $\tilde{U} = 8$ is used for comparison with
results of Ref. \onlinecite{Yamaji}, and b) an infinite $\tilde{U}$
is used to see if the attraction induced by EPI, alone, is enough
for superconductivity.

We examine the d-wave $\Delta_{\bf k} = \Delta(\cos k_x - \cos
k_y)$, extended s-wave $\Delta_{\bf k} = \Delta(\cos k_x + \cos
k_y)$ and s-wave $\Delta_{\bf k} = \Delta$ order parameters.
Following previous work \cite{Yamaji} periodic boundary conditions
are used in one direction and anti-periodic boundary conditions are
used in the other.  The value of $\lambda$ is varied to study the
effect of increasing EPI and we find the value of $\Delta$ at which
the energy minimum occurs. We also measure the energy of clustered
states to ensure our results are stable against the formation of
immobile clusters. In our results all energies quoted are \emph{per
electron}; note that Yamaji et al. \cite{Yamaji} use energy per
site.

\begin{figure}
\includegraphics[width = 63mm, angle=-90]{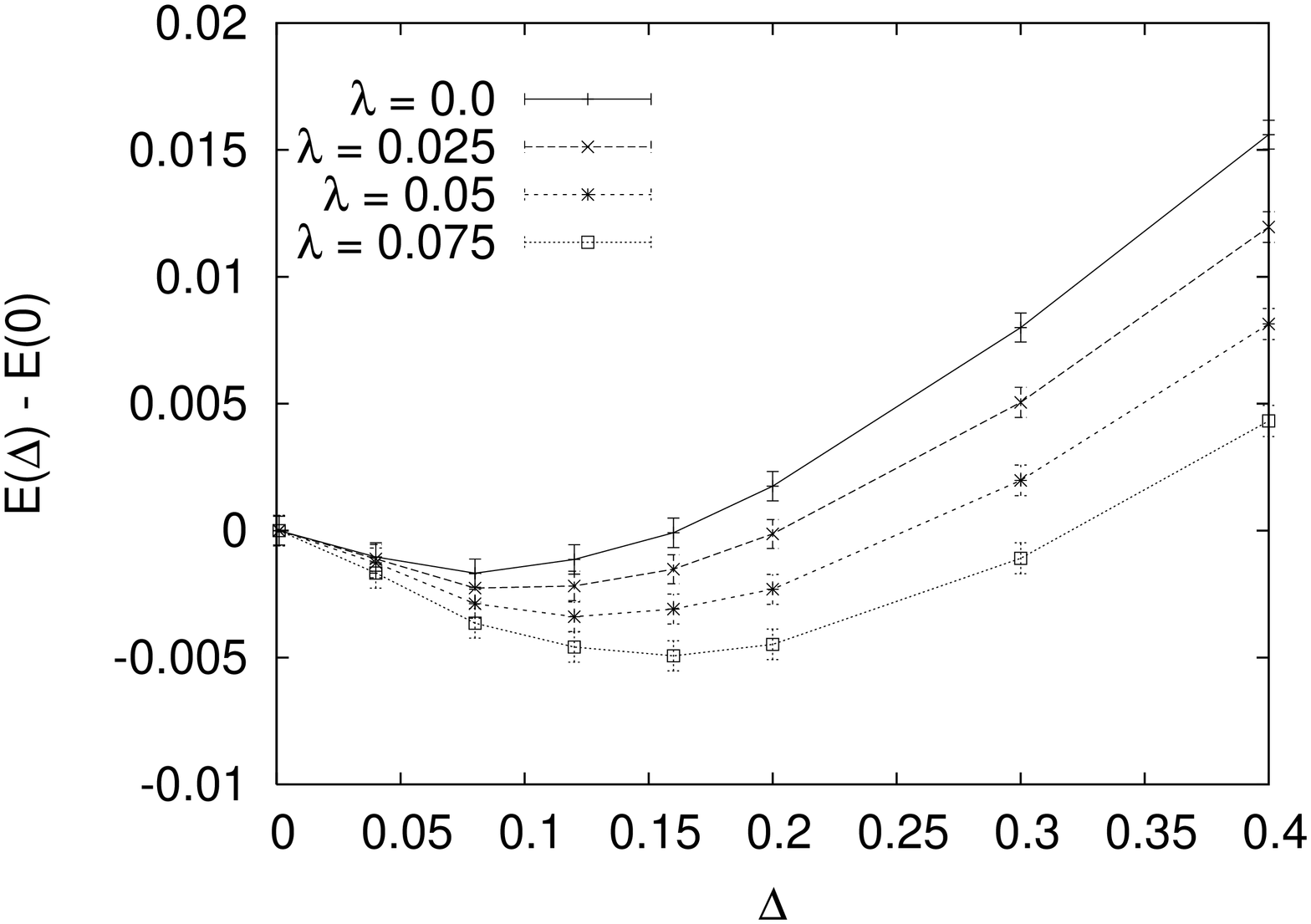}
\includegraphics[width = 63mm, angle=-90]{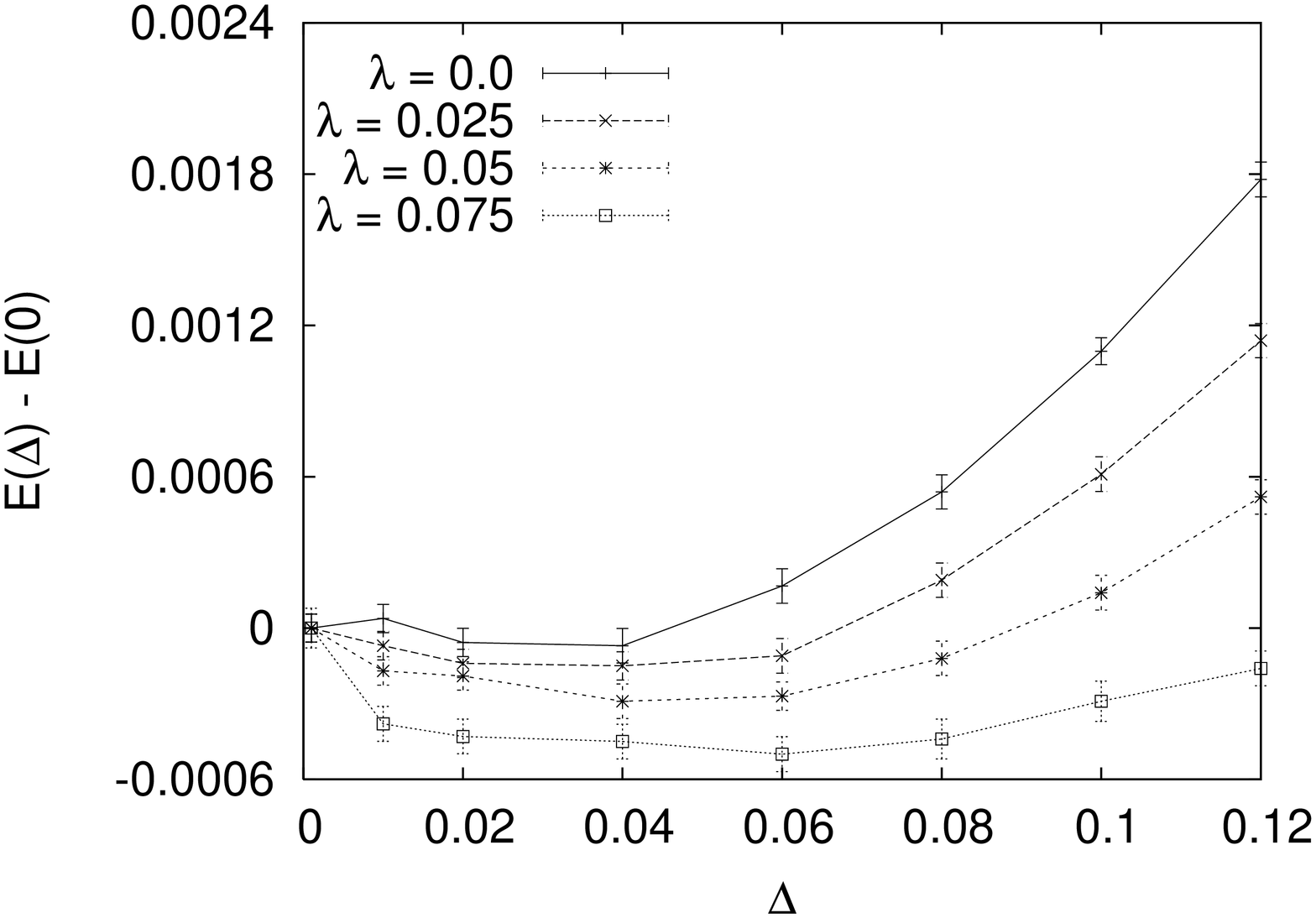}
\caption{Condensation energy per electron (in units of $\tilde{t}$)
versus the amplitude of the superconducting d-wave order-parameter
for $\tilde{U}=8$, top, and $\tilde{U}=\infty $, bottom,  with
different EPI coupling, $\lambda$.} \label{DWave}
\end{figure}

In the following, all energies are quoted in units of $\tilde{t}$.
For the $\lambda = 0$ and $\tilde{U} = 8$ case shown in
Fig.\ref{DWave} (top), we recover Yamaji's result~\cite{Yamaji} with
a minimum at $\Delta \approx 0.08$ and energy per electron of
$(-0.8780 \pm 0.0004)$, the normal state energy is $(-0.8763 \pm
0.0004)$. It can clearly be seen that the effect of increasing EPI
is to increase the depth of the minimum and therefore the stability
of the superconducting state. Our minimum energy per electron is
$(-4.4008 \pm 0.0004)$ at $\Delta = 0.16$ and $\lambda=0.075$.  The
maximum condensation energy gain $|E(\Delta) - E(0)|$ with $\lambda
= 0$ is $(0.0017 \pm 0.0006)$  and $(0.0049 \pm 0.0006)$ per
electron with $\lambda = 0.075$.

A surprising result is found for infinite Coulomb repulsion, where
mapping to the $t$--$J$ model would result in $|J|=4t^2/U=0$ (so no
spin fluctuations are present). Fig.\ref{DWave} (bottom) shows an
increasing condensation energy gain with $\lambda$ for $\tilde{U} =
\infty$: the maximum condensation energy gain is $(0.00050 \pm
0.00007)$ per electron at $\Delta = 0.06$. While this energy gain is
not large enough to guarantee stability against other trial wave
functions, it does suggest a d-wave superconducting state exists for
$\tilde{U} = \infty$ with strong enough $\lambda$, that is $d$-wave
order can be induced without spin-fluctuations (as expected, the
$\lambda=0$ curve shows no superconducting order).

\begin{figure}
\includegraphics[width = 63mm, angle=-90]{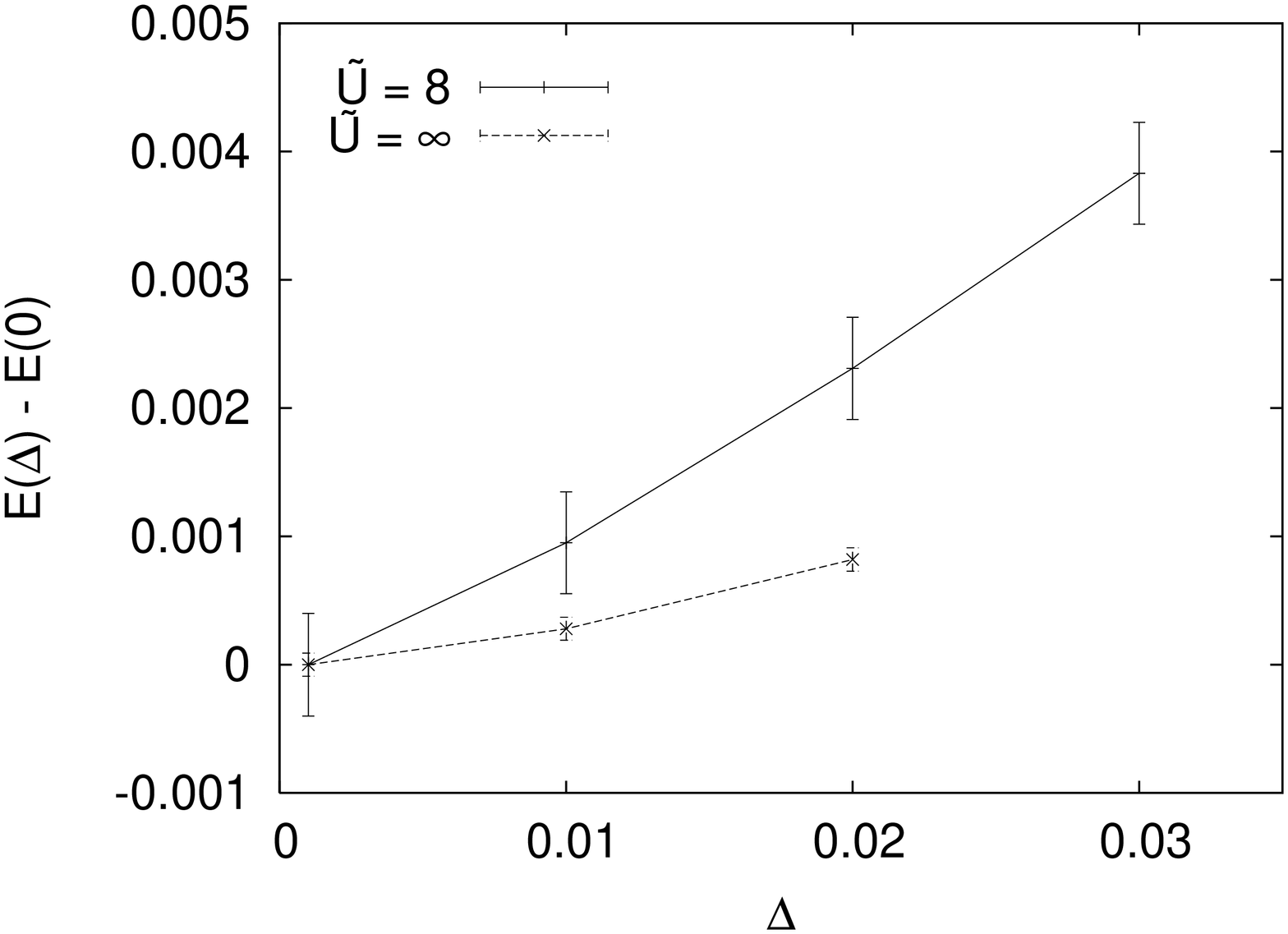}
\includegraphics[width = 63mm, angle=-90]{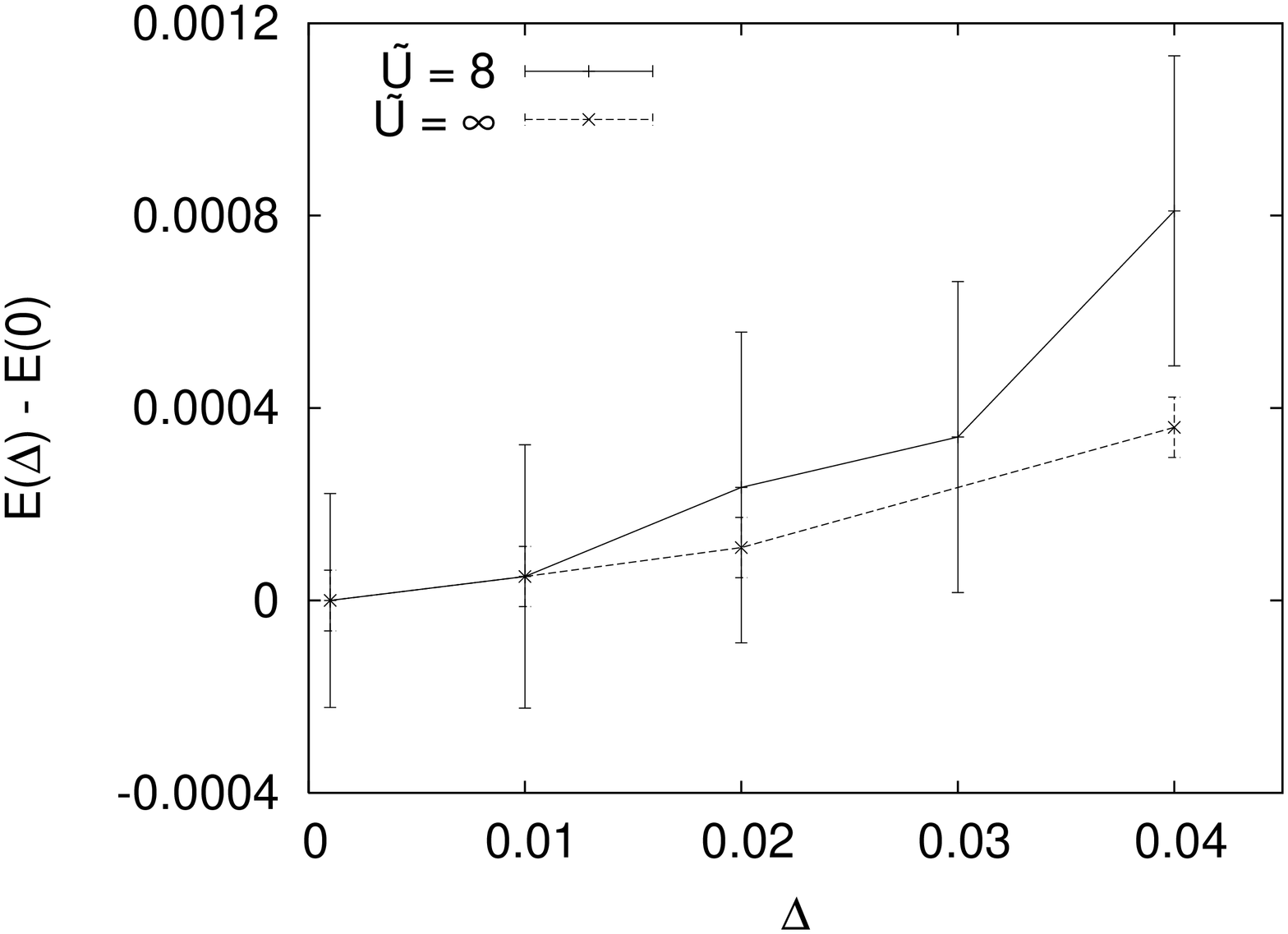}
\caption{Condensation energy per electron (in units of $\tilde{t}$)
versus the amplitude of the superconducting s-wave order-parameter
(top) for
 $\tilde{U} = \infty $ and $\tilde{U} = 8$,  showing no s-wave ground state.  The condensation energy  versus
the amplitude of the extended s-wave order-parameter (bottom) for
 $\tilde{U} = \infty $ and $\tilde{U} = 8$, showing no extended s-wave ground state. }
\label{SWaveUInf}
\end{figure}

We examined the energies of various cluster formations.  The lowest
clustering energy for $\lambda = 0.075$ is $-3.6824$ per electron,
so that the system is stable against the formation of clusters. It
is found that there is no s-wave or extended s-wave superconducting
state for both $\tilde{U} = 8$ and $\tilde{U} = \infty $ with
$\lambda$ up to $0.075$ see Fig.\ref{SWaveUInf}.

Accurate QMC studies have proven that the Hubbard $U$ and $t$--$J$
models are unlikely to account for the high critical temperatures of
cuprate superconductors \cite{Imada}. On the other hand ab-initio
LDA calculations of EPI have led to a conclusion that phonons fail
to explain high-temperature superconductivity either \cite{manske}.
It is well known that LDA badly underestimates the role of the
Coulomb correlations, incorrectly predicting that the parent
compounds of cuprate superconductors (such as La$_2$CuO$_4$) are
metallic rather than insulating. Also, the anisotropy  of the
electron response functions determined with LDA is much smaller than
the experimentally observed value in these layered materials. It is
not surprising at all that EPI turns out to be very weak in this
``metallic'' picture due to electron screening of the long-range
electron-ion interaction. Moreover, it has been noted that the
inclusion of Hubbard $U$ via the LDA$+U$ scheme significantly
enhances the EPI strength \cite{zha} since the system becomes a
doped Mott insulator with poor screening in at least the c-axis
direction as anticipated in Refs. \onlinecite{alemot} and
\onlinecite{asa}.

These theoretical results, as well as many experiments
\cite{zhao,lanzara, mic1,ita,tal,ega,rez}, tell us that any realistic
theory of high-temperature cuprate superconductors should include
strong electron-electron correlations and the finite-range EPI on
equal footing, as previously discussed by Alexandrov and Mott
\cite{alemot}. When an extended Fr\"{o}hlich EPI is considered, the
picture changes drastically compared with the local Holstein
interaction, where carriers are readily self-trapped on a single
lattice site even by a moderate electron-phonon coupling
\cite{polarons}. It is interesting to note the profound effect of long
range electron-phonon coupling on the magnitude of the superconducting
order, since such long range interactions have equally dramatic
effects when electron-phonon coupling is strong. For example, we found
that small mobile bipolarons could exist when electrons interact via a
strong \emph{finite-range} Fr\"{o}hlich EPI with high-frequency
optical phonons \cite{asa,alekor2,jim2}.

Here, we showed using the VMC method, that even a relatively weak
Fr\"{o}hlich EPI is sufficient to induce a d-wave superconducting
state with substantial condensation energy in a doped Mott-Hubbard
insulator and/or strongly-correlated metals. The calculated
condensation energy and the difference between clustering and
superconducting energies in a wide region of $\lambda$ are larger
than any difference between VMC and the more exact GBMC algorithm of
Aimi and Imada~\cite{Imada}, so that our superconducting state is
very likely to be robust with respect to a more complicated choice
of the trial function. As a result we believe that the true origin
of high-temperature superconductivity lies in the proper combination
of strong electron-electron correlations with long range
electron-phonon interactions.

We would like  to thank  Pavel Kornilovitch  for long-standing
collaboration and helpful discussions. We acknowledge support of
this work by EPSRC (UK) (grant number EP/C518365/1).

\bibliography{TomFinal}

\end{document}